# A Study of Spin-Flipping in Sputtered IrMn using Py-based Exchange-Biased Spin-Valves.

R. Acharyya, H.Y.T. Nguyen, W.P. Pratt Jr., and J. Bass.

Department of Physics and Astronomy, Michigan State University, East Lansing, MI 48823.

Abstract

To study spin flipping within the antiferromagnet IrMn, we extended prior Current-Perpendicular-to-Plane (CPP) Giant Magnetoresistance (GMR) studies of Py-based exchange-biased-spin-valves containing IrMn inserts to thicker IrMn layers—5 nm ≤ $t_{IrMn}$ ≤ 30 nm. Unexpectedly, A$\Delta$R= A($R_{AP} - R_P$)--the difference in specific resistance between the anti-parallel (AP) and parallel (P) magnetic states of the two Py layers—did not decrease with increasing $t_{IrMn}$, for $t_{IrMn}$ ≥ 5 nm, but rather became constant to within our measuring uncertainty. This constant looks to be due mostly to a new, small MR in thin Py layers. The constant complicates isolating the spin-diffusion length, $l_{sf}^{IrMn}$, in bulk IrMn, but $l_{sf}^{IrMn}$ is probably short, ≤ 1 nm. Similar results were found with FeMn.

I. Introduction.

Antiferromagnetic (AF) metals, like IrMn and FeMn, find wide use in spintronics. In Giant Magnetoresistance (GMR), they are used to make spin valves, where the AF pins an adjacent ferromagnetic (F) layer such as Co, Py, or Ni [1-3]. They are also a crucial part of studies of antiferromagnetic GMR (AFGMR) [4]. Especially for studies in the Current-Perpendicular-to-Plane (CPP) geometry, it is important to know their transport properties, including spin flipping within AFs and at AF/F and AF/N (N = non-magnet) interfaces.

We previously studied spin flipping in IrMn up to thickness $t_{IrMn}$= 5nm [5]. Our data showed that spin flipping at IrMn/Cu interfaces is strong, but raised the possibility that spin-flipping within the IrMn might be weaker. To clarify the situation, in this paper we extend our studies to $t_{IrMn}$ = 30 nm.

II. Samples and Techniques.

Our sample fabrication and measuring techniques are described in [6,7]. The present samples are exchange-biased spin-valves (EBSVs) of the form FeMn(8)/Py(24)/Cu(10)/IrMn($t_{IrMn}$)/Cu(10)/Py(24), with all thicknesses in nm. The AF FeMn layer pins the adjacent Permalloy (Py) layer by exchange bias. The pinned Py is exchange decoupled from the free Py layer by 20nm of Cu. Initially a large, negative, in-plane magnetic field, H = - 200 to - 400 Oe, aligns the magnetizations of the two Py layers parallel to each other (P-state). As shown in Fig. 1 of ref. [5], when the field direction is reversed to H ≥ + 10 to + 20 Oe, the free layer magnetization reverses to give antiparallel (AP) order of the Py magnetizations. Finally, at H ≥ + 180 to + 230 Oe, the pinned layer reverses, returning to P order.

To achieve uniform current flow in the CPP geometry, the samples are sandwiched between crossed, 1.1 mm wide, 150 nm thick, Nb strips, which become superconducting at our measuring temperature of 4.2K (see inset to Fig. 1). The quantity of interest is the difference in specific resistances (AR = Area × Resistance) between the AP and P states , A$\Delta$R = AR(AP) – AR(P), where A is the overlap area of the Nb strips.

For $t_{IrMn}$ ≤ 1nm, we were able to map out in detail the transitions from P to AP to P states [5]. For $t_{IrMn}$ ≥ 1.5 nm, however, the signal for individual A$\Delta$R measurements decreased to near noise. To increase the signal to noise, we measured A$\Delta$R for such samples 100 times at each of three selected fields: - 300 Oe (P-state); +50 Oe (AP state); and + 300 Oe (P-state), then repeated the measurements in the same order, and averaged the 400 measurements at - 300 Oe and + 300 Oe, and the 200 measurements at + 50 Oe, and took the difference as A$\Delta$R [5].

III. Theory.

From Valet-Fert (VF) theory of CPP-MR [5,8-10], as the thickness $t_{IrMn}$ of the IrMn insert grows, A$\Delta$R should decrease as in Eq.(1), where the IrMn/Cu interface is assumed to have finite thickness, $t_I$ ~ 0.6-0.8 nm [11].

A$\Delta$R = [Kexp(- spin flip)]/(AR$_o$+AR$_{IrMn}$+2AR$_{IrMn/Cu}$)  (1)

Here K is a proportionality constant, AR$_0$ is the specific resistance of the sample excluding the IrMn insert and the IrMn/Cu interfaces, AR$_{IrMn}$ is due to the IrMn bulk, and 2AR$_{IrMn/Cu}$ is due to the two IrMn/Cu interfaces. The spin-flip exponent has contributions from both the IrMn/Cu interfaces and the bulk of the IrMn [4,7,8]. As the interface forms, A$\Delta$R decreases for two reasons: (a), growth of the 2AR$_{IrMn/Cu}$ term in the denominator, and (b) increase in spin-flipping from the interfaces in the exponent in the numerator. After the interfaces are formed, A$\Delta$R decreases further due to the same two terms in bulk IrMn: (a) growth of AR$_{IrMn}$ (= $\rho_{IrMn}t_{IrMn}$, where $\rho_{IrMn}$ = 1260 n$\Omega$m) in the denominator, and (b) growth of spin-flipping in the numerator. However, once $t_{IrMn}$ > $l_{sf}^{IrMn}$, the bulk IrMn term in the denominator saturates at AR$_{IrMn}$ = $\rho_{IrMn}$x $l_{sf}^{IrMn}$. Thereafter, the decay of A$\Delta$R is dominated by the term – $t_{IrMn}/l_{sf}^{IrMn}$ due to bulk spin-flipping within IrMn. In principle, $l_{sf}^{IrMn}$ can be determined from the slope of a plot of log (AΔR) vs $t_{IrMn}$ in this thick $t_{IrMn}$ limit.

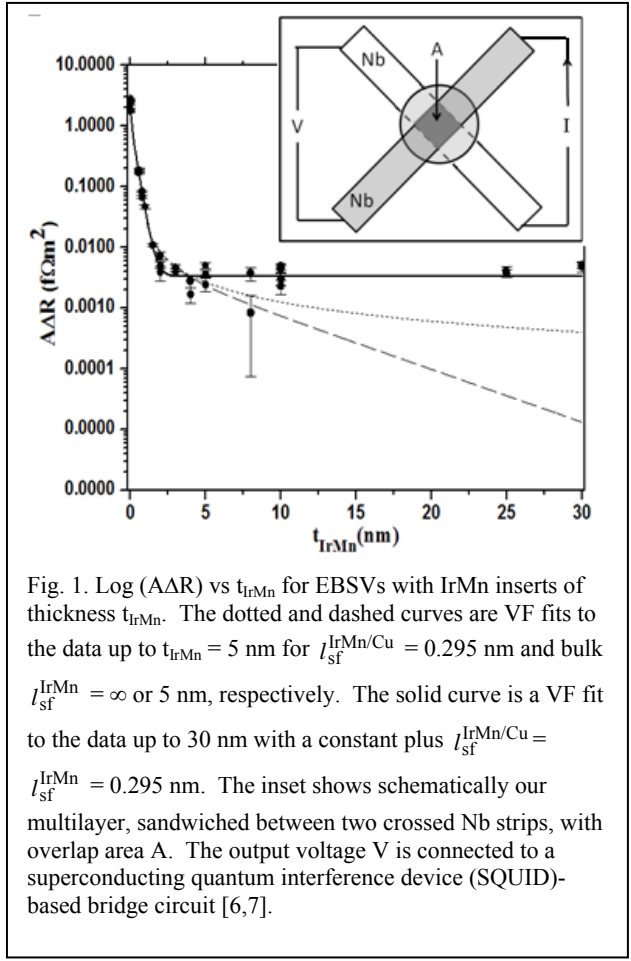

Fig. 1. Log (AΔR) vs $t_{IrMn}$ for EBSVs with IrMn inserts of thickness $t_{IrMn}$. The dotted and dashed curves are VF fits to the data up to $t_{IrMn}$ = 5 nm for $l_{sf}^{IrMn/Cu}$ = 0.295 nm and bulk $l_{sf}^{IrMn}$ = ∞ or 5 nm, respectively. The solid curve is a VF fit to the data up to 30 nm with a constant plus $l_{sf}^{IrMn/Cu}$ = $l_{sf}^{IrMn}$ = 0.295 nm. The inset shows schematically our multilayer, sandwiched between two crossed Nb strips, with overlap area A. The output voltage V is connected to a superconducting quantum interference device (SQUID)-based bridge circuit [6,7].

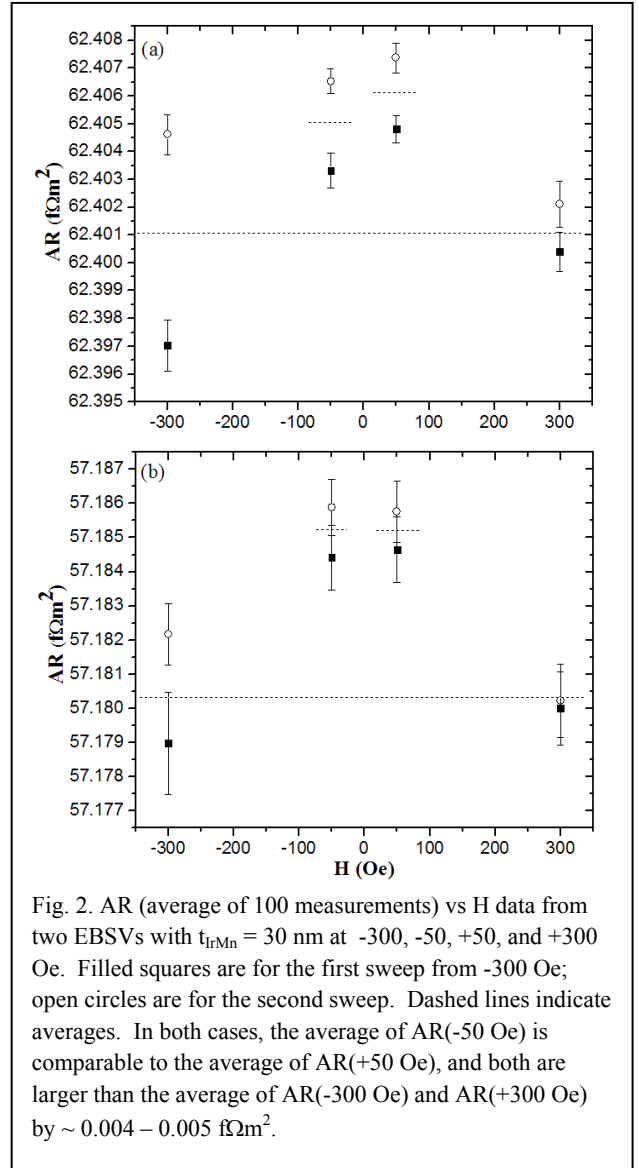

Fig. 2. AR (average of 100 measurements) vs H data from two EBSVs with $t_{IrMn}$ = 30 nm at -300, -50, +50, and +300 Oe. Filled squares are for the first sweep from -300 Oe; open circles are for the second sweep. Dashed lines indicate averages. In both cases, the average of AR(-50 Oe) is comparable to the average of AR(+50 Oe), and both are larger than the average of AR(-300 Oe) and AR(+300 Oe) by ~ 0.004 – 0.005 fΩm².

IV. Data and Analysis.

As shown in Fig. 1, in a prior study up to $t_{IrMn}$ = 5 nm [3], AΔR decreased logarithmically by about a factor of 50 up to $t_{IrMn}$ ~ 1nm, comparable to the IrMn/Cu interface thickness. The slope of this line gave an effective spin-diffusion length, $l_{sf}^{IrMn/Cu}$ ≈ 0.24 nm, which we ascribed to strong spin-flipping at the IrMn/Cu interface. For 1.5 nm ≤ $t_{IrMn}$ ≤ 5 nm, the rate of decrease slowed. If this slowing was due to weaker spin-flipping in the bulk IrMn, the data did not extend far enough to constrain the spin-diffusion length $l_{sf}^{IrMn}$ in the bulk IrMn. As examples, for the present paper we used VF theory to predict the behavior beyond $t_{IrMn}$ = 5 nm for $l_{sf}^{IrMn}$ = ∞ (dotted curve in Fig. 1), and $l_{sf}^{IrMn}$ = 5 nm (dashed curve in Fig. 1). This more rigorous analysis increased $l_{sf}^{IrMn/Cu}$ slightly to 0.295 nm. Up to $t_{IrMn}$ = 5 nm, the dotted and dashed curves in Fig. 1 are nearly indistinguishable. To better understand spin flipping in IrMn, we clearly needed to extend measurements to $t_{IrMn}$ >> 5nm. The present study goes out to $t_{IrMn}$ = 30nm.

These extended results are also shown in Fig. 1. Instead of a slower decay, as illustrated by the dotted and dashed curves, AΔR for $t_{IrMn}$ ≥ 5 nm becomes constant to within our measuring uncertainty, at a value AΔR = 0.0037 ± 0.0002 fΩm². The solid curve in Fig.1 shows a VF fit to all of the data combining this constant background with assumed $l_{sf}$ = 0.295 nm for both the IrMn/Cu interface and bulk IrMn (this VF $l_{sf}^{IrMn/Cu}$ is slightly different than the simple exponential fit assumed above and in ref. [5]). We'll argue below that this constant term is due to Py rather than to IrMn. If so, then only a few data points in the range $t_{IrMn}$ = 2-5 nm remain to be associated with 'bulk IrMn'. The lack of any obvious excess in AΔR over the constant for $t_{IrMn}$ ≥ 3 nm suggests that any exponential contribution from bulk

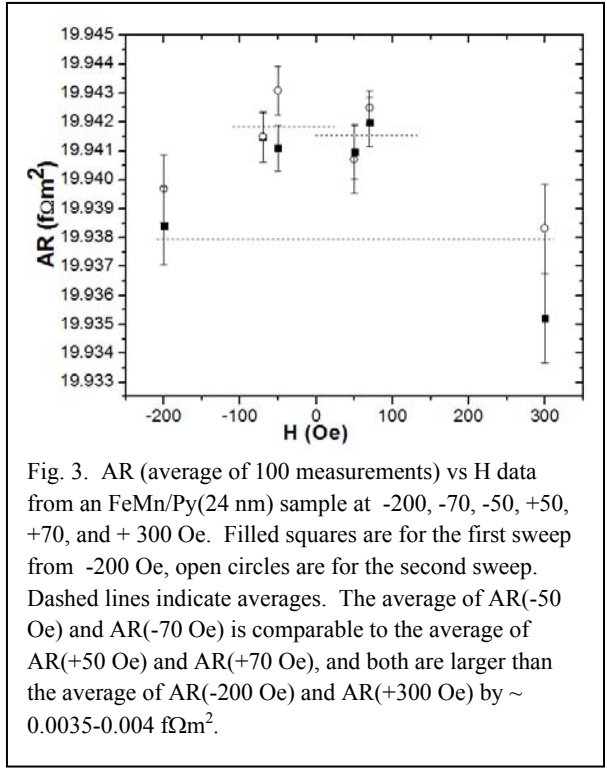

Fig. 3. AR (average of 100 measurements) vs H data from an FeMn/Py(24 nm) sample at -200, -70, -50, +50, +70, and +300 Oe. Filled squares are for the first sweep from -200 Oe, open circles are for the second sweep. Dashed lines indicate averages. The average of AR(-50 Oe) and AR(-70 Oe) is comparable to the average of AR(+50 Oe) and AR(+70 Oe), and both are larger than the average of AR(-200 Oe) and AR(+300 Oe) by ~ 0.0035-0.004 f$\Omega$m$^2$.

IrMn must decay much faster than the dashed curve. Uncertainty in the magnitude of the constant term limits our ability to set a tight bound on $l_{sf}^{IrMn}$, but it looks to be very short, most likely ~ 1 nm or less.

As a first test that the constant might not be related to a CPP-GMR, we remeasured several samples with $t_{IrMn} \geq 1.5$ nm, including now a measurement at -50 Oe between the -300 Oe and +50 Oe ones At – 50 Oe, the sample should still be in the P state. So a value of A$\Delta$R resulting from a larger AR at – 50 Oe than that at – 300 Oe should not be a CPP-GMR. Fig. 2 shows examples of field sweeps for two samples with $t_{IrMn}$ = 30 nm. In both cases, AR at – 50 Oe is larger than at – 300 Oe by ~ 0.004 -0.005 f$\Omega$m$^2$, and this difference is comparable to the increase at + 50 Oe.

If A$\Delta$R for $t_{IrMn} \geq 4$ nm in Fig. 1 is not a CPP-GMR, then what is it? To see if it might be associated with just the Py layers, we made samples with the following structures (and Nb electrodes on both sides): (a) Cu(5)/Py(t)/Cu(5) with t = 50 nm; (b) FeMn(8)/Py(t) or Py(t)/FeMn(8) with t = 50, 24, and 12 nm, some with and some without field pinning. Fig. 3 shows AR vs H for a FeMn(8)/Py(24) sample at fields -200, -70, -50, +50, + 70, and +300 Oe. The average ARs at – 70 Oe and – 50 Oe are larger than those at - 200 Oe and + 300 Oe, and comparable to those at + 50 Oe and + 70 Oe. The average A$\Delta$R was not sensitive to pinning but, intriguingly, grew modestly with decreasing t, a behavior not yet understood. The overall average for – 50 Oe of all these samples was A$\Delta$R(Py) = 0.003 ± 0.001 f$\Omega$m$^2$.

Finally, we tried to reduce spin flipping at the top IrMn/Cu interface, by inserting 1 to 5nm of Nb or Ru between the IrMn and Cu. Whereas we expect Mn in Cu to have a magnetic moment, Kondo effect studies [12] suggest no such moment in Nb or Ru. On average, such inserts gave little or no change in A$\Delta$R suggesting that the strong spin-flipping at IrMn/Cu interfaces does not result mainly from "loose" Mn moments in the Cu.

Lastly, we also extended equivalent measurements to $t_{FeMn}$ = 30 nm for the other widely used AF, FeMn. The results obtained were similar to those for IrMn.

V. Summary and Conclusions.

To summarize, we extended our prior studies [5] of the spin-flipping properties of IrMn to thicker IrMn layers—i.e., $t_{IrMn}$ extending from 5 nm to 30 nm. Instead of A$\Delta$R decaying with increasing $t_{IrMn}$, as expected for a finite $l_{sf}^{IrMn}$ in bulk IrMn, we found an approximately constant value of A$\Delta$R = 0.0037 ± 0.0002 f$\Omega$m$^2$ for $t_{IrMn} \geq 4$ nm. Studies of Py layer inserts showed that most of this constant signal was unrelated to GMR, but originates from a new field dependence of AR for Py. We concluded that spin-flipping is significant in the IrMn/Cu interface and that 'loose' Mn moments in Cu are not the source of this strong spin flipping. The presence of the constant term complicates estimation of the spin diffusion length of bulk IrMn, but it is most likely short-- $\leq$ 1 nm. Similar results were also found for FeMn.

Acknowledgments: This work was supported in part by NSF grant DMR 08-04126 and a grant from the Korea Institute for Science and Technology (KIST).